\documentclass[conference]{IEEEtran}
\IEEEoverridecommandlockouts
\usepackage{cite}
\usepackage{amsmath,amssymb,amsfonts}
\usepackage{algorithmic}
\usepackage{graphicx}
\usepackage{textcomp}
\usepackage{xcolor}
\usepackage{placeins}
\usepackage{hyperref}
\def\BibTeX{{\rm B\kern-.05em{\sc i\kern-.025em b}\kern-.08em
    T\kern-.1667em\lower.7ex\hbox{E}\kern-.125emX}}
\begin{document}
\setlength{\parskip}{0pt}
\setlength{\parindent}{1em}
\title{Natural Language Interface for Firewall Configuration\\

}

\author{
\IEEEauthorblockN{Farid Taghiyev}
\IEEEauthorblockA{\textit{Tandon School of Engineering} \\
\textit{Department of Electrical and Computer Engineering} \\
\textit{New York University}\\
New York, US \\
ft2337@nyu.edu}
\and
\IEEEauthorblockN{Ali Aslanbayli}
\IEEEauthorblockA{\textit{Tandon School of Engineering} \\
\textit{Department of Electrical and Computer Engineering} \\
\textit{New York University}\\
New York, US \\
aa12947@nyu.edu}

}

\maketitle

\begin{abstract}
This paper presents the design and prototype implementation of a natural language interface for configuring enterprise firewalls. The framework allows administrators to express access control policies in plain language, which are then translated into vendor specific configurations. A compact schema bound intermediate representation separates human intent from device syntax and in the current prototype compiles to Palo Alto PAN OS command line configuration while remaining extensible to other platforms. Large language models are used only as assistive parsers that generate typed intermediate representation objects, while compilation and enforcement remain deterministic. The prototype integrates three validation layers, namely a static linter that checks structural and vendor specific constraints, a safety gate that blocks overly permissive rules such as any to any allows, and a Batfish based simulator that validates configuration syntax and referential integrity against a synthetic device model. The paper describes the architecture, implementation, and test methodology on synthetic network context datasets and discusses how this approach can evolve into a scalable auditable and human centered workflow for firewall policy management.
\end{abstract}

\begin{IEEEkeywords}
firewall configuration, natural language interface, large language models, network security, policy verification
\end{IEEEkeywords}

\section*{GitHub Repository}
\url{https://github.com/ftaghiyev/firewall-configuration-interface}

\section{Introduction}

In a typical enterprise, a single firewall platform is rarely sufficient. Most environments operate a combination of devices, such as Palo Alto PA-series, Fortinet FortiGate, and Cisco Firepower, each using distinct configuration languages and semantics. As a result, security teams must manually translate high-level human policies like "allow HR to access the payroll SaaS over HTTPS" or "block outbound SSH from guests" into vendor-specific rules. The correctness of these configurations depends on fine-grained details such as object naming, zone definitions, rule order, application signatures, and system defaults. Although the underlying logic of most rules can be represented by a simple five-tuple (source IP, destination IP, source port, destination port, and protocol), maintaining consistent and non-conflicting configurations across different devices remains complex and error-prone. Studies have shown that misconfigurations are common and often correlate with the size and complexity of rule sets, redundant or shadowed entries, and unclear user interfaces which is expected from the inherently low-level nature of current configuration tools \cite{b1}.

This report studies a natural language interface that lets an administrator express the policy directly in plain text and then compiles it into the correct device configuration for a chosen target platform. Formally, the problem is to map (X), a natural-language description of what flows should be allowed or denied, into (Y), a concrete configuration for a particular firewall family. Large language models (LLMs) are a natural fit for the front end because they can extract entities and resolve ambiguous phrasing. Modern LLMs are built on the Transformer architecture, which introduced the self-attention mechanism to capture contextual relationships between tokens \cite{b13}. This capability allows models to interpret dependencies such as actor, action, and condition in policy descriptions, making them suitable for translating intent into structured firewall rules. The contextual reasoning strength demonstrated in few-shot learning frameworks further supports their use in interpreting unstructured administrative requests when coupled with schema-constrained outputs \cite{b14}. However, we should not rely on a general model to emit device commands directly. The safer approach is to have the model produce a structured, typed intermediate representation (IR) that captures the five-tuple and necessary context, and only then compile that IR deterministically to each vendor's syntax. This separation gives us a place to validate the intent, check for conflicts, and run simple checks before anything touches production [2, 6].

There is some evidence that this direction is viable. Studies on chatbot-generated rules for firewalls and IDSs indicate that generic prompting leads to low success rates, but task specialization and training significantly improve correctness, which suggests that a domain-constrained model can act as a useful code generator when we wrap it in validation \cite{b2}. At the same time, work on LLM-assisted network defense, like DDoS mitigation planning, shows practical prompting patterns (role conditioning, schema-aware inputs, external tools) that we can borrow for translating policies rather than classifying traffic \cite{b7}. On the other side, detection-focused systems such as ML-based web application firewalls and newer LLM-aided intrusion detection are useful for context, but they target runtime traffic classification, not control-plane configuration. In this project, the goal is to compile least-privilege access rules from human intent, not to detect attacks online [4, 8].

Safety is central: we do not want an LLM to have direct access to devices or to produce unconstrained text that may be accepted blindly. Recent guidance on securing LLM agents reframes classic principles such as least privilege, complete mediation, fail-safe defaults for model-driven systems and argues for clear capability boundaries, typed outputs, and interposition layers that filter inputs and outputs \cite{b9}. In our setting, that translates into a mediated LLM layer that only produces IR objects obeying a schema, followed by a compiler and an automated verification harness. An additional content-policy layer, similar in spirit to an "LLM firewall", can screen the prompts and the proposed rules for unsafe patterns such as any-to-any permits, broad wildcard objects, or missing directions, which helps keep the interface usable without sacrificing guardrails \cite{b5}.

A structured exchange format makes the above much simpler. YANG-style models have already been used to exchange configuration fragments with LLMs in lab settings, and results show that accuracy is high when the schema is explicit and the tasks are bounded \cite{b6}. We adopt that idea and define a compact vendor-agnostic IR that extends the five-tuple with just enough metadata to compile to Palo Alto, Fortinet, and Cisco Firepower. The LLM’s job is then to read the natural-language request, resolve references to known enterprise objects, and propose a minimal set of allow or deny entries in that IR. Constrained decoding and schema validation reduce hallucinations after which vendor back ends take care of ordering, object creation, and platform-specific details. Finally, a Batfish-based simulator validates the generated configuration for syntax and referential integrity against a synthetic device header before any change plan is produced [1, 2, 6, 10, 11].

To place this work among related efforts, we will refer to three neighboring lines of research. First, the firewall usability literature motivates moving the interface closer to how administrators actually think and write policies, rather than forcing them into device-specific rule editors \cite{b1}. Second, results on LLM-driven defenses suggest practical input representations and prompting styles that make low-level networking tasks more reliable, which we reuse in our translation pipeline [7, 10]. Third, rule-generation loops from the application layer, for example automatic WAF hardening against SQL injection with generate-and-retest workflows, illustrate how to keep a model on track by validating every change against a test corpus. We adapt that mindset to network reachability tests instead of attack corpora \cite{b3}.

In short, the contribution is a design for a natural language to multi-vendor configuration compiler that keeps the LLM in a constrained role. The interface accepts policy statements in plain English, converts them into a typed IR, validates and compiles them into vendor-specific rule sets, and verifies the result before staging. Throughout, we follow recent security recommendations for model-based systems, so the model proposes changes but does not apply them, and every step is logged for auditability [5, 9, 12]. The next sections will detail the IR, the prompting and retrieval strategy, the vendor back ends, and the verification and safety layers, and will discuss limitations such as ambiguity in user requests, uneven feature parity across vendors, and domain drift as networks evolve.

\section{Proposed Solution}

\subsection{Architecture and Intermediate Representation}

\begin{figure*}[t]
\centerline{\includegraphics[width=\textwidth]{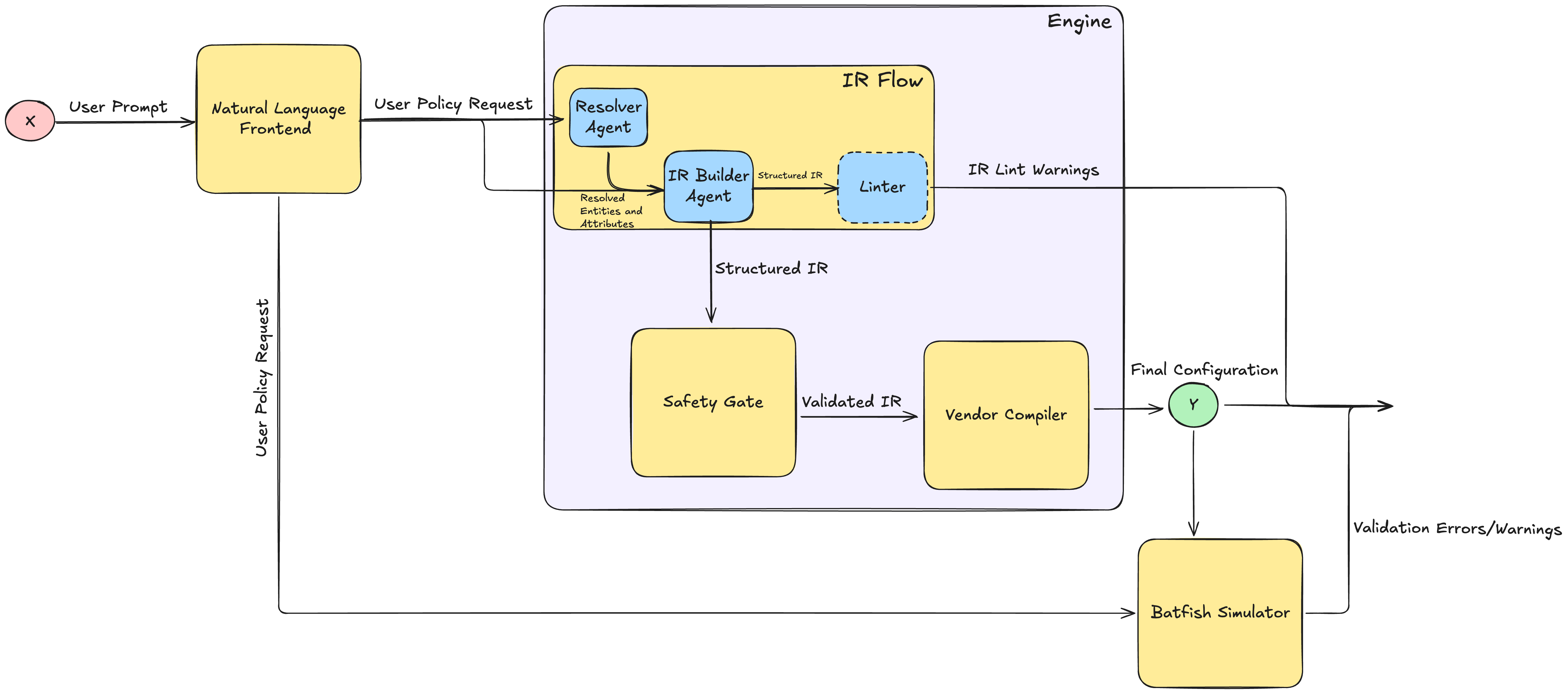}}
\caption{Proposed system architecture}
\label{fig:architecture}
\end{figure*}

The system follows a left-to-right pipeline in which a natural-language front end turns the user’s message and selected network context into a structured policy request, and the engine converts that request into device-specific configuration (see Fig.~\ref{fig:architecture}). Inside the engine, an \emph{IR Flow} sub-pipeline first runs a Resolver Agent, then an IR Builder Agent, and finally a linter before handing the result to downstream safety and compilation stages. The Resolver Agent grounds free-form names in the prompt to concrete objects, zones, and services that exist in the network context, while the IR Builder Agent assembles the canonical rule structure in the intermediate representation (IR). The central design choice is to separate policy intent from device syntax via this compact, typed IR. Each IR rule captures the five-tuple plus the minimal context needed for safe compilation: action, protocol, source and destination (as objects, subnets, or user groups), optional ports or application labels, ingress/egress zones, direction, priority, logging, and time windows. A YANG-style discipline is applied so that all fields remain explicit and machine-checkable, aligning with prior research indicating that schema-bound exchanges with LLMs improve accuracy and validation efficiency \cite{b6}. This structured separation helps mitigate common usability and configuration errors in real-world rule sets, including shadowing, redundancy, and unintended reachability, while removing the need for administrators to work directly in vendor-specific configuration languages \cite{b1}.

\subsection{Natural-Language Front End, Resolution, and Compilation}
The natural-language front end accepts the user's prompt together with a selected network context file that describes objects, zones, and services, and forwards both as a policy request to the engine (Fig.~\ref{fig:architecture}). Within the IR Flow block, two schema-bound LLM agents perform intent extraction. The Resolver Agent receives the raw text plus context and maps ambiguous phrases such as "HR laptops" or "Payroll SaaS" to concrete entities defined in the context (address objects, zones, and services). Its output is a shallow, typed intent record that already normalizes actions, directions, and protocol families. The IR Builder Agent then consumes this resolved intent and constructs one or more rules in the canonical IR, filling all mandatory fields and attaching metadata such as the raw policy string and any ambiguities the model detected. Both agents are invoked with role-conditioned prompts and constrained decoding so that they may only emit JSON conforming to the IR schema, following prior results that schema-bounded exchanges with LLMs substantially improve validation and correctness \cite{b2,b6}.

After the IR has been produced, downstream processing is deterministic. In the current prototype we implement a single vendor back end for Palo Alto PAN-OS. The vendor compiler reads the validated IR and generates CLI fragments that create or reuse address, service, and schedule objects and insert security rules with explicit source and destination zones. The translation is intentionally one-way and side-effect free: the compiler never calls devices directly and either succeeds with a reproducible configuration snippet or produces a structured error describing which IR field could not be mapped to the target platform. The modular structure of the compiler allows additional back ends (for example, Fortinet or Cisco Firepower) to be added without changing the IR or the front-end agents.

\subsection{Verification, Safety, and Governance}
The pipeline applies four verification and analysis steps that act on the intermediate representation and the compiled configuration. These steps are the general linter, the vendor specific linter, the Safety Gate, and a Batfish based simulator. The first two are advisory checks that never stop execution. The Safety Gate is a hard gate that blocks unsafe rules from reaching the compiler. Batfish simulation runs after compilation and provides additional feedback about the resulting configuration and its synthetic environment.

The first layer is the general IR linter. This component checks for structural issues that are independent of the target platform. It detects empty source or destination lists, duplicate rule identifiers, invalid port ranges, misuse of ports under ICMP or protocol ``any'', invalid priority values, and unsupported actions. These checks match the behavior implemented in the prototype and define the baseline integrity conditions for IR rules. The general linter only produces warnings and the pipeline continues even if warnings are present.

The second layer is the vendor specific linter for Palo Alto PAN OS. This linter evaluates platform constraints and hygiene rules that are specific to PAN OS. It verifies that the protocol belongs to a supported set, confirms the presence of source and destination zones, and checks that schedule names contain only valid characters. It flags unusual constructs such as schedules on deny rules and identifies cases where custom service objects are required because no built in PAN OS application covers the given protocol and port combination. It also warns when inbound rules originate from trust zones or when outbound rules target trust zones and when the same object appears in both source and destination fields. This behavior corresponds to the logic implemented in the vendor linter and, again, only produces warnings. The compiled configuration is still generated even if vendor linter warnings exist.

The third layer is the Safety Gate which enforces high level security constraints. The Safety Gate inspects the final IR and returns a boolean flag together with a list of errors. It rejects any rule that allows traffic from any source to any destination and flags rules that omit source or destination zones, that have empty source or destination lists, or that lack a protocol specification. If any such error is present, the Safety Gate marks the IR as unsafe and the pipeline does not invoke the vendor compiler or Batfish. In the current prototype, this component is the only verification layer that can stop further execution and therefore acts as a hard guardrail against clearly unsafe or incomplete rules.

The final layer is Batfish based simulation. After a configuration passes the Safety Gate and is compiled into PAN OS CLI, the Batfish manager wraps the generated commands in a synthetic device header that defines basic interfaces, a default virtual router, and a minimal set of zones and address objects derived from the network context. It then initializes a Batfish snapshot and runs a sequence of analyses that check parsing issues, undefined references, and unused structures. The results are returned as a list of warnings and errors with severities. In the current prototype, these results are surfaced to the operator for inspection but do not block the pipeline. Batfish is therefore used as an external consistency and hygiene check over the compiled configuration and the constructed environment rather than as a hard gate. Together, these four components provide structural validation, platform specific analysis, explicit safety enforcement, and offline simulation before any configuration is considered ready for review.

\subsection{Illustrative Walk-Through}
Consider a prompt like "Allow Finance to reach Vendor-Invoices over HTTPS on weekdays 08:00--18:00 and block outbound SMTP from Guests." The front end extracts two rules and the Resolver Agent maps names via retrieval: Finance to an AD group or subnet, Vendor-Invoices to FQDNs or an application object, and Guests to a guest zone and subnet. The IR Builder Agent records TCP/443 with a time window for the allow, and TCP/25 to untrusted networks for the deny, plus logging and explicit zones. The Palo Alto back end generates or reuses address and application objects, defines a schedule, and inserts deny-then-allow rules in the security rulebase. The linter and Safety Gate enforce structural and security constraints, for example rejecting an empty source list or an accidental any-to-any allow. Finally, the Batfish simulator confirms that the PAN-OS CLI is syntactically valid and that all referenced objects and services exist in the synthetic device model. If Batfish or the safety layers raise issues, the system surfaces them along with the IR so that the operator can either rephrase the request or modify the context. Extending this final stage to execute reachability queries over the Batfish snapshot is conceptually straightforward and is left as future work.

\section{Limitations}

Natural-language requests are expressive but often underspecified. Even with a five-tuple backbone, everyday phrasing tends to omit direction, zones, or precise object names. "Payroll" might be a SaaS domain set, an internal subnet, or an application signature, and "HR laptops" could map to a user group, a dynamic address group, or both. The typed IR forces these fields to be explicit, however correctness still hinges on a clean source of truth implying that if the enterprise catalog is stale or inconsistent, the compiler will produce rules that are syntactically valid yet semantically off target. On the other hand, heterogeneous devices add friction because features do not align perfectly across vendors and an application-aware allow on one platform can degrade to a port-based rule on another, or layered policies may have no direct counterpart. The back end either rejects such cases with a clear diagnostic or emits a conservative approximation, which preserves safety but can create path asymmetries that must be surfaced during review. Robustness and privacy are handled through strict constraints on model behavior and data handling. The LLM operates only within a schema-defined intermediate representation and has no direct access to devices or live configurations. However, natural-language inputs can still contain ambiguous or adversarial phrasing that might unintentionally produce overly permissive rules. To prevent this, an I/O policy layer intercepts unsafe requests and blocks risky patterns before processing. All name resolution occurs locally, and any sensitive identifiers are redacted in prompts. Every interaction is logged with audit trails to maintain transparency without exposing raw network information.

Correctness at compile time may also degrade over time. Networks drift as DNS for SaaS endpoints changes, subnets are reallocated, and vendor application signatures evolve. The combination of linting, the Safety Gate, and Batfish-based simulation helps by re-validating syntax and basic consistency against the current context and by surfacing hygiene issues such as missing objects or overly broad scopes, but these safeguards only hold if we re-validate periodically and version object mappings so past approvals remain reproducible. In practice, the limitations are manageable when we pair technical controls with routine hygiene by keeping the object catalog accurate, requiring explicit fields in requests, documenting cross-vendor differences in the change plan, prefer fail-safe defaults, and scheduling lightweight regression checks to catch drift early. Prior work largely reinforces this stance since operator error tends to grow with rule complexity, which motivates a small, explicit IR, and domain-constrained prompting or training improves accuracy only when outputs are validated after generation [1, 2].

\section{Evaluation Methodology}

\begin{figure}[h]
\centerline{\includegraphics[width=0.5\textwidth]{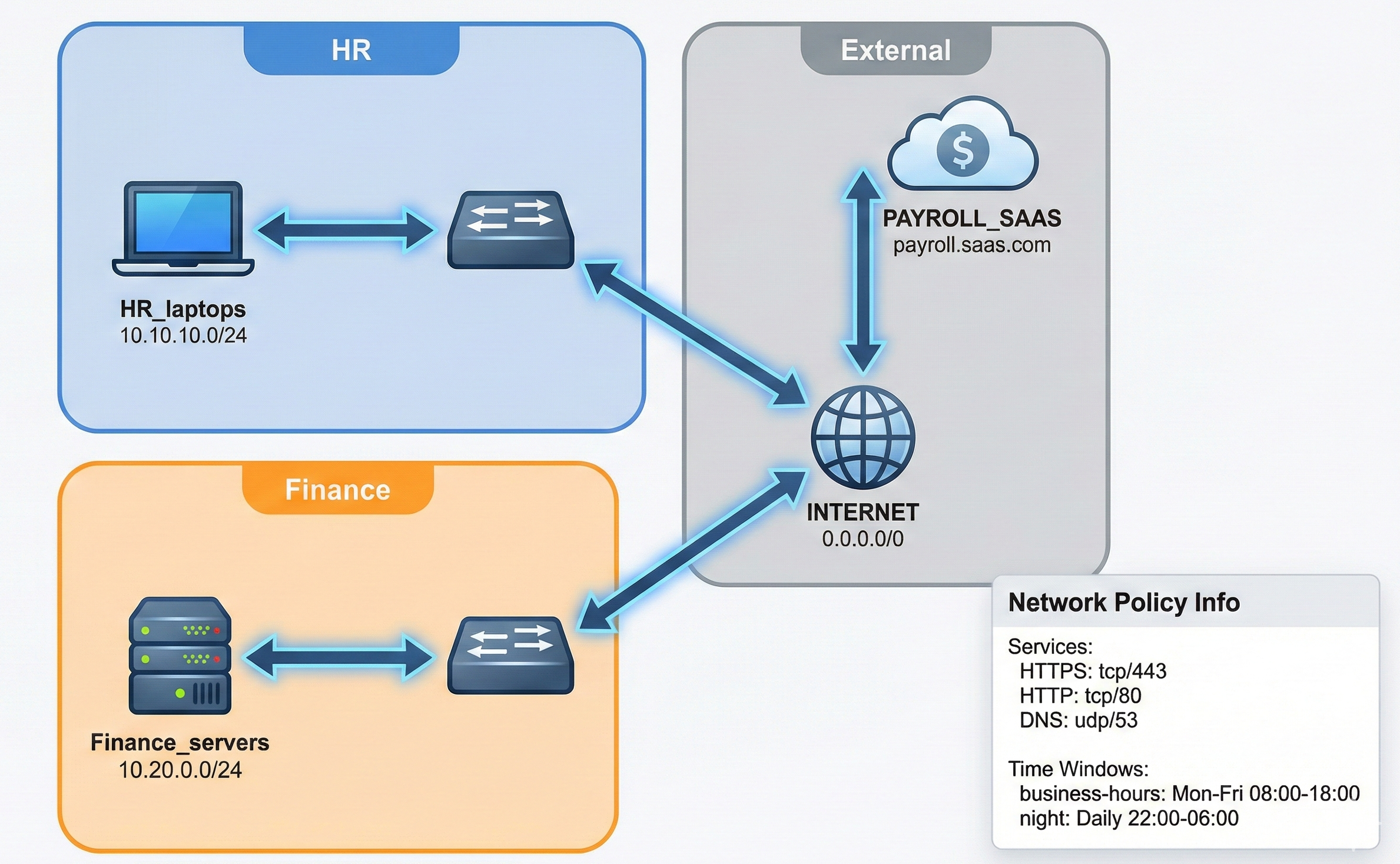}}
\caption{Example Network Configuration}
\label{fig:network-conf}
\end{figure}

The prototype is evaluated offline using synthetic network configuration datasets rather than live devices. Each dataset is a JSON file that describes a small enterprise, industrial, or cloud deployment in terms of address objects, security zones, and named services. Together these files play the role of a network context that the Resolver Agent and compiler use to ground natural-language policy intents. Example contexts include an e-commerce platform and a smart factory, each containing multiple zones, server and client objects, and a set of basic services such as HTTP, HTTPS, and DNS. See a sample from our dataset in Fig.~\ref{fig:network-conf}

Testing is organized in terms of \emph{triplets}. A single test case consists of three elements: (i) a natural-language query that acts as the user’s policy request (for example, "Allow WebServer to reach DB on TCP 5432 during business hours"), (ii) an expected IR instance that encodes the canonical, vendor-agnostic rule the system should produce, and (iii) an expected piece of PAN-OS CLI that represents the correct vendor-specific configuration for that rule. The natural-language input and network context are fed into the full pipeline, and the resulting IR and CLI are compared to the reference artifacts.

Evaluation proceeds in two layers. The first layer measures \emph{semantic accuracy} of the LLM-based agents by checking whether the generated IR exactly matches the expected IR for each triplet. Because the IR is fully typed and schema-bound, this check reduces to structural equality on the JSON representation, capturing errors such as swapped zones, missing services, or incorrect actions. The second layer measures \emph{syntax accuracy} of the deterministic compiler by comparing the generated CLI with the reference CLI using a text-similarity metric (for example, Python's \texttt{difflib.SequenceMatcher}). The current prototype uses a strict threshold of 100\% similarity to count a test as passed, which reflects the security-sensitive nature of firewall configuration.

Under this evaluation, the system achieves a pass rate of approximately 85\% over a collection of hand-crafted triplets. Most remaining failures are attributable to subtle ambiguities in the natural-language requests or to edge cases in PAN-OS syntax that are not yet captured by the linters and compiler templates. Usability and end-to-end operational safety in real networks are left for future work and would require controlled user studies and long-running shadow deployments, respectively.

\section{Prototype}

\begin{figure*}[!t]
\centering
\includegraphics[width=\textwidth]{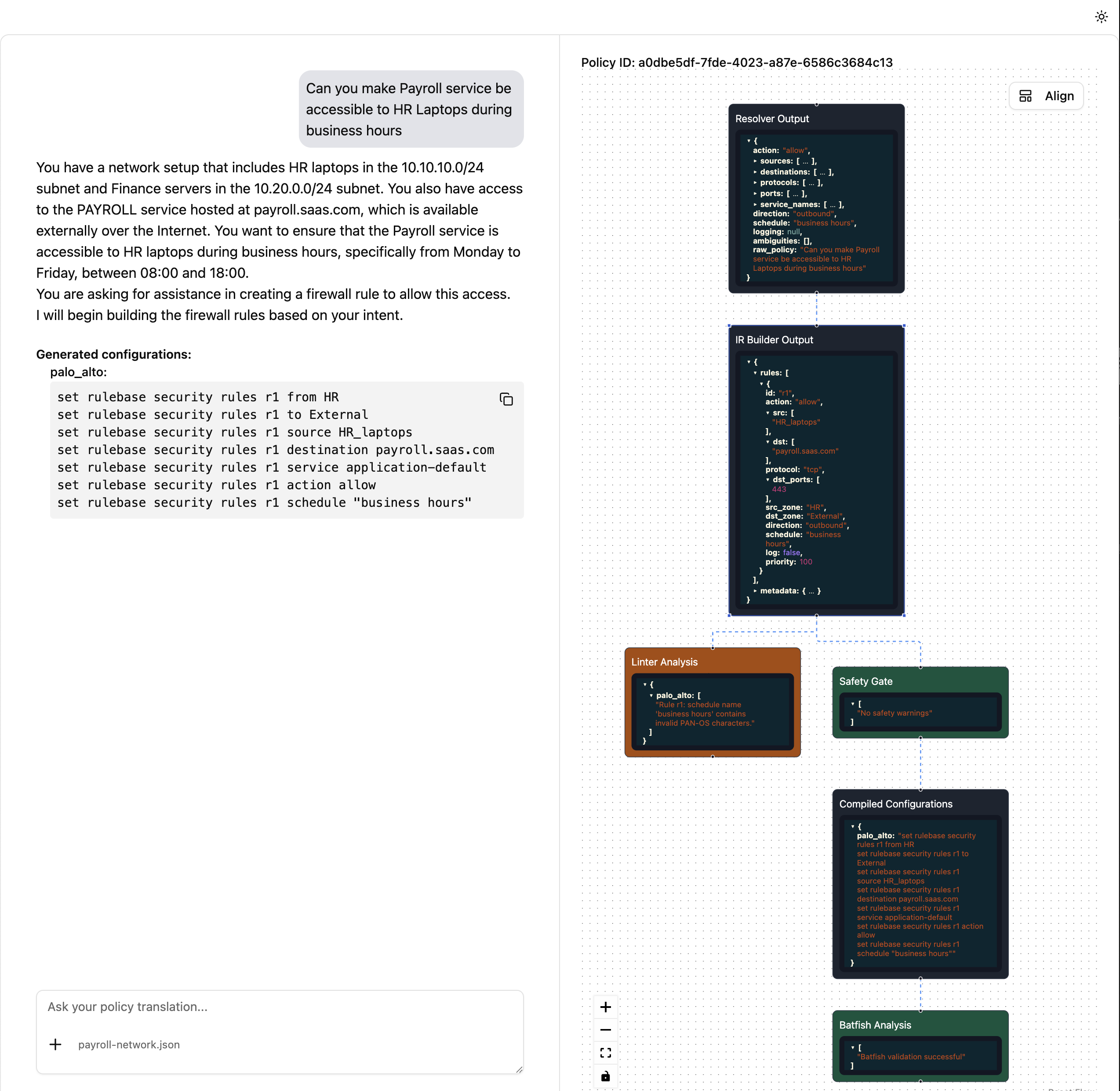}
\caption{Working Prototype Demonstration}
\label{fig:prototype}
\end{figure*}

A minimal working prototype was developed to demonstrate the end to end interaction of the system. The user interface is a simple chat window where an operator can upload a network context as a JSON file or paste the same information directly as plain text. Once the context is provided, the operator can describe a policy in natural language and the system processes the request through the full pipeline. A screenshot of the prototype is shown in Fig.~\ref{fig:prototype}.

After the request completes successfully, the interface displays an interactive diagram on the right side. This visualization is implemented using ReactFlow and it mirrors the pipeline described in the system architecture. Each node corresponds to one stage in the processing flow. These stages include the Resolver Agent, the IR Builder Agent, the general linter and the Palo Alto linter, the Safety Gate, the vendor compiler, and the Batfish simulator. Each node presents the structured input and output produced at that step which includes resolved entities, the intermediate representation, linter warnings, Safety Gate decisions, compiled configuration lines, and Batfish validation messages. The diagram also highlights the fact that linter warnings do not stop execution, that the Safety Gate can block unsafe IR before compilation, and that Batfish provides offline analysis of the compiled configuration. This visualization makes the internal decision process transparent and helps operators understand how the system interpreted their request.

The prototype does not deploy configurations to real devices. Its primary purpose is to validate the usability of the natural language interface and to illustrate how structured processing and layered validation increase safety.

\section{Conclusion and Future Work}

This report presented a complete design and prototype of a natural language interface for translating high level security policies into vendor specific firewall configurations. The system separates intent extraction from device level syntax through a schema bound intermediate representation. The new architecture introduces a clear sequence of components that process a policy request. The Resolver Agent grounds names to objects in the network context and the IR Builder Agent constructs a typed rule set that captures the full semantics of the request. Downstream stages apply structural linting, high level safety checks, deterministic compilation, and Batfish based validation. The verification layers have clearly defined roles. The general linter and the Palo Alto linter provide non stopping feedback about structural and platform specific issues so the operator can see inconsistencies without interrupting execution. The Safety Gate is the only blocking component and it stops the pipeline when a rule is unsafe or incomplete at the IR level. Batfish then evaluates the compiled configuration in a synthetic environment and reports parsing or reference issues without affecting control flow. This layered design keeps the language model in a constrained role and maintains a transparent and auditable workflow that reduces configuration errors and prevents unsafe outputs from reaching deployment.

The prototype demonstrates how these components operate together in practice. The chat based interface lets operators upload a context and describe a policy in plain language. The resulting execution trace is visualized and it shows the output of each stage in the pipeline. This helps users understand how their request was interpreted and why a specific configuration was produced. The evaluation using synthetic network contexts and triplets confirms that schema bound intent extraction combined with deterministic compilation can achieve high accuracy in producing both intermediate representation objects and vendor specific configuration lines. 

There are several promising directions for future work. One important extension involves adding full reachability analysis to the Batfish stage so that the system can verify intended and unintended flows rather than only checking syntax and object references. Another direction involves support for multiple back ends including Fortinet and Cisco Firepower so that the same intermediate representation can target heterogeneous environments. The system can also benefit from retrieval augmented prompts that adapt to organizational naming conventions and policy libraries. Further improvements include continuous context validation so that the system can detect drift or outdated objects and warn the operator before compilation. The prototype can also be expanded with guided disambiguation steps where the interface asks clarifying questions whenever a policy is underspecified. Finally, large scale testing with real operators would help refine the user experience and measure how well the natural language interface reduces configuration time and error rates.

The long term vision is a unified policy workflow that accepts human intent in natural language, verifies correctness through layered validation, and produces vendor specific configurations in a predictable and safe manner. The results of this work show that such a workflow is feasible and that structured intermediate representations combined with constrained model outputs provide a practical foundation for future network configuration systems.

\end{document}